\newif\ifastroph

\astrophtrue

\documentclass[preprint2]{aastex6} % or aastex?

\newcommand{\myemail}{mulders@lpl.arizona.edu}

\slugcomment{Manuscript Accepted in AJ}

\shorttitle{Metallicity of short period exoplanets}
\shortauthors{Mulders et al.}

\usepackage{natbib,graphicx,xspace,amsmath}
\newcommand{\Teff}{\ensuremath{T_{\rm eff}}\xspace}
\newcommand{\FeH}{\ensuremath{{\rm [Fe/H]}}\xspace}
\newcommand{\DeltaFeH}{\ensuremath{{\Delta \rm [Fe/H]}}\xspace}
\newcommand{\FeHcount}{\ensuremath{{\rm \overline{[Fe/H]}}_{\rm KOI}}\xspace}
\newcommand{\FeHocc}{\ensuremath{{\rm \overline{[Fe/H]}}}\xspace}
\newcommand{\logg}{\ensuremath{{\rm log}~g}\xspace}
\newcommand{\dex}{\ensuremath{{\rm dex}}\xspace}
\newcommand{\KOI}{planet candidate\xspace}
\newcommand{\KOIs}{planet candidates\xspace}
\newcommand{\focc}{\ensuremath{f_{\rm occ}}\xspace}

\begin{document}

\title{A Super-Solar Metallicity For Stars With Hot Rocky Exoplanets}

\author{Gijs D. Mulders\altaffilmark{1}, Ilaria Pascucci\altaffilmark{1}, and D\'aniel Apai\altaffilmark{1,2}}
\affil{Lunar and Planetary Laboratory, The University of Arizona, Tucson, AZ 85721, USA}
\email{\myemail}
\author{Antonio Frasca}
\affil{INAF -- Osservatorio Astrofisico di Catania, via S. Sofia, 78, 95123 Catania, Italy}
\author{Joanna Molenda-\.Zakowicz\altaffilmark{3}}
\affil{Astronomical Institute, University of Wroc{\l}aw, ul.\,Kopernika 11, 51-622 Wroc{\l}aw, Poland}

% additional affiliations
\altaffiltext{1}{Earths in Other Solar Systems Team, NASA Nexus for Exoplanet System Science}
\altaffiltext{2}{Department of Astronomy, The University of Arizona, Tucson, AZ 85721, USA}
\altaffiltext{3}{Department of Astronomy, New Mexico State University, Las Cruces, NM 88003, USA}

\begin{abstract}
The host star metallicity provide a measure of the conditions in protoplanetary disks at the time of planet formation.
Using a sample of over 20,000 \textit{Kepler} stars with spectroscopic metallicities from the \texttt{LAMOST} survey, we explore how the exoplanet population depends on host star metallicity as a function of orbital period and planet size. 
We find that exoplanets with orbital periods less than 10 days are preferentially found around metal-rich stars ([Fe/H]$\simeq{} 0.15\pm0.05$ dex). The occurrence rates of these hot exoplanets increases to $\sim{} 30\%$ for super-solar metallicity stars
from $\sim{} 10\%$ for stars with a sub-solar metallicity. Cooler exoplanets, that reside at longer orbital periods and constitute the bulk of the exoplanet population with an occurrence rate of $\gtrsim{} 90\%$, have host-star metallicities consistent with solar.
At short orbital periods, $P<10$ days, the difference in host star metallicity is largest for hot rocky planets ($<1.7 ~R_\oplus$), where the metallicity difference is [Fe/H]$\simeq 0.25\pm0.07$ dex. 
The excess of hot rocky planets around metal-rich stars implies they either share a formation mechanism with hot Jupiters, or trace a planet trap at the protoplanetary disk inner edge which is metallicity-dependent.
We  do not find statistically significant evidence for a previously identified trend that small planets toward the habitable zone are preferentially found around low-metallicity stars.  Refuting or confirming this trend requires a larger sample of spectroscopic metallicities.
\end{abstract}


\section{Introduction}
Stellar metallicity is a good proxy of the initial metallicity of the protoplanetary disks, which in turn has an important impact on planet formation. Together with the disk mass, the disk metallicity determines the amount of solids available in protoplanetary disks for planet formation.
Higher mass stars host more massive disks \citep[e.g.][]{2013ApJ...771..129A,2016arXiv160803621P} and a larger metallicity corresponds to a larger amount of condensible solids in the disk. Therefore, higher stellar masses and metallicities result in more building blocks available in the disk for planet formation. For gas giant planets, a correlation between planet occurrence and stellar metallicity \citep{2000A&A...363..228S, 2010PASP..122..905J,2012Natur.486..375B, 2013A&A...551A.112M} and stellar mass \citep{2007ApJ...670..833J,2010PASP..122..905J,2015A&A...574A.116R} has been well established. Theoretically, this can be understood as massive cores need to reach a critical mass of $\sim 10 ~M_\oplus$ to undergo runaway gas accretion before the gas dissipates, which is more likely to occur in disks with more solids \citep[e.g.][]{2004ApJ...616..567I,Alibert:2011kg,2012ApJ...751...81J,2012A&A...541A..97M}.

\begin{figure}
	\includegraphics[width=\linewidth]{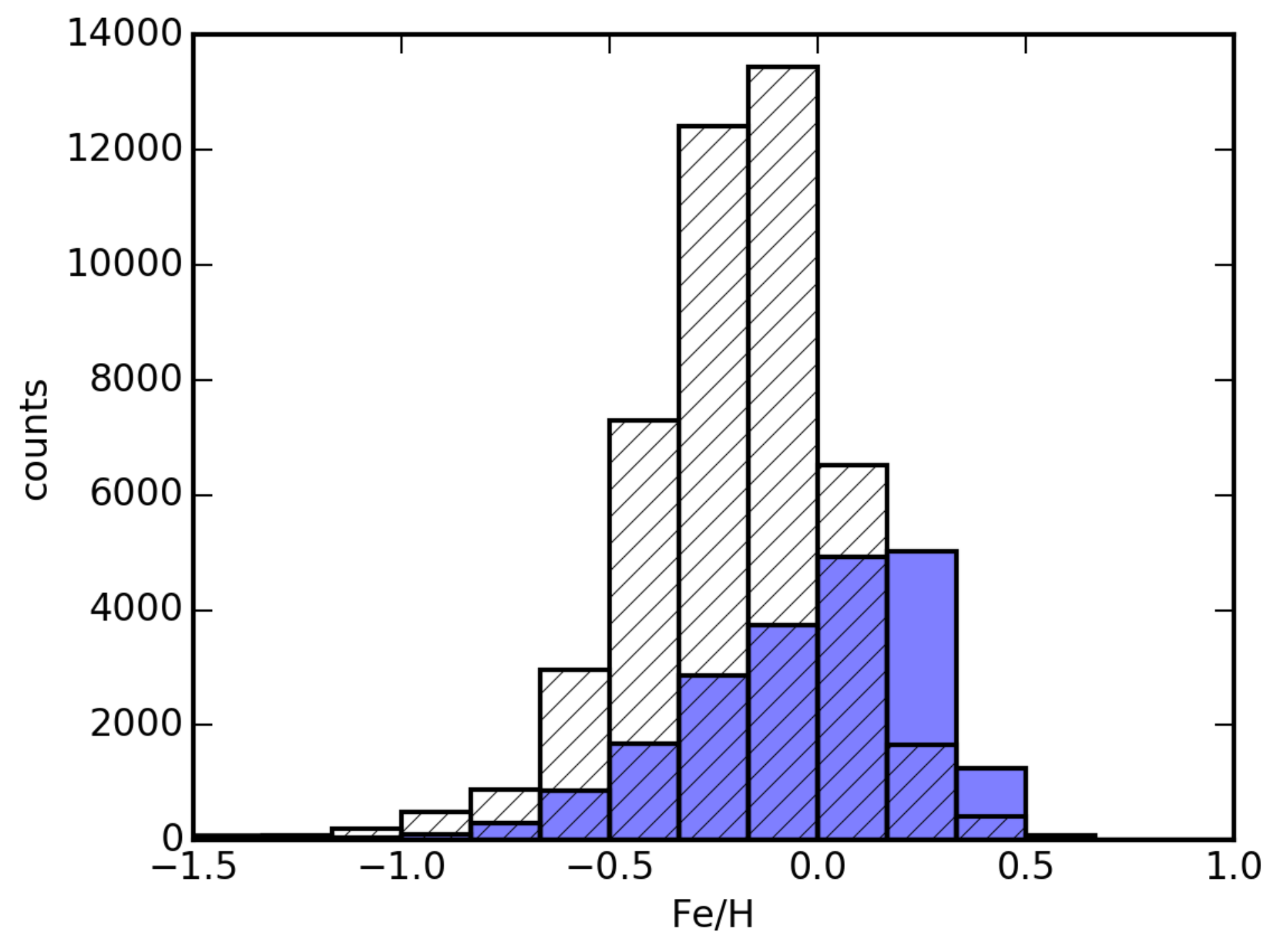}
	\caption{Histogram of spectroscopic metallicities of main-sequence stars in the \textit{Kepler} field from \texttt{LAMOST} ( blue). Photometric metallicities from \cite{2014ApJS..211....2H} for a sample of \textit{Kepler} targets brighter than Kpmag$=14$, representative of the brightness limit of the \texttt{LAMOST} stellar sample, are shown for comparison  (hatched). 
	\label{f:hist}
	}
\end{figure}

For smaller planets, those that have been found in abundance with the \textit{Kepler} spacecraft, correlations between planet occurrence and host star mass and metallicity are different from those for giant planets, and are less straightforward to interpret. The occurrence rate of these planets is \textit{anti}-correlated with stellar mass \citep{2012ApJS..201...15H,2015ApJ...798..112M}. This indicates a \textit{larger} amount of solids forming planets around low-mass stars \citep{2015ApJ...814..130M}, at least at short ($\lesssim ~1$ yr) orbital periods, in contrast with observed protoplanetary disk dust masses \citep{2013ApJ...773..168M,2013ApJ...771..129A,2016ApJ...827..142B,2016arXiv160405719A,2016arXiv160803621P}. The correlation between stellar metallicity and planet occurrence rate disappears towards lower mass planets, indicating that these planets can form around stars with a wide range of metallicities \citep{2008A&A...487..373S, 2012Natur.486..375B}. The large number of transiting planets with spectroscopically determined metallicities indicate only small rocky planets ($\lesssim ~1.7 R_\odot$) show no correlation with stellar metallicity, while larger mini-Neptunes ($1.7-3.9 R_\odot$) show a correlation with metallicity but one that is weaker than for giant planets \citep{2014Natur.509..593B,2015ApJ...808..187B}, however see \cite{2015ApJ...799L..26S}.  A metallicity correlation for mini-Neptunes is also observed in a sample with measured planet masses \citep{2016MNRAS.461.1841C}. Another potential diagnostic of the planet formation process is the dependence of planet orbital period on host star metallicity. Different studies have pointed out underpopulated regions in the host star metallicity-orbital period diagram for small planets, at various orbital periods ranging from 5 to 70 days \citep{2013ApJ...763...12B,2013A&A...560A..51A,2015MNRAS.453.1471D,2016OLEB..tmp...30A}.

In this paper, we revisit these results in the context of the exoplanet population. We use a dataset of over 20,000 medium-resolution spectroscopic metallicities for \textit{Kepler} target \textit{stars} from the \texttt{LAMOST}-\textit{Kepler} project \citep{2015ApJS..220...19D,2016arXiv160609149F}. This large dataset provides a homogeneous planet survey with a well-characterized detection bias, enabling us to estimate survey completeness. \textit{For the first time, we are able to calculate planet occurrence rates based on spectroscopically determined stellar metallicities.} We describe the target sample and methodology in section \ref{s:analysis} and present the main results of the metallicity dependence of the planet population on orbital period in section \ref{s:results}. We evaluate a potential trend towards the habitable zone in \S \ref{s:HZ}, and discuss potential origins for the excess of hot rocky planets around high-metallicity stars in section \ref{s:D}. We summarize our results and present and outlook for future research in section \ref{s:C}.

\begin{table}
	\title{Planet Occurrence Rates and Host Star Metallicities}
	\centering
   \begin{tabular}{l l l l l | l}\hline\hline
   KOI & $R_P$ & $P$   & $f_{\rm occ}$ & \FeH  & $f_{\rm occ}$ \\
       & [cm]  & [day] &               & [dex] &   \\
   \hline
   K00001.01 & 7.9e+09 & 2.5 & 2.9e-04 & 0.28 & 5.6e-04 \\
   K00005.01 & 3.8e+09 & 4.8 & 4.8e-04 & 0.36 & 9.1e-04 \\
   ... & ... & ... & ... & ... & ... \\
   K06242.03 & 8.9e+08 & 78.9 & 6.5e-03 & -0.41 & 1.5e-02 \\
   K06246.01 & 1.0e+09 & 9.1 & 8.3e-04 & 0.32 & 1.6e-03 \\
   \hline\hline\end{tabular}
	\par % end centering?
	\caption{Planet Occurrence Rates and Host Star Metallicities.
	The final column denotes the occurrence rate of the planet in the super-solar ($\FeH \ge 0$) or sub-solar ($\FeH < 0$) metallicity sample.
	Table \ref{t:occ} is published in its entirety in the electronic edition of the Astrophysical Journal. A portion is shown here for guidance regarding its form and content.}
	\label{t:occ}
\end{table} 

\section{Analysis}\label{s:analysis}
\subsection{Metallicities}\label{s:LAMOST}
The observing strategy, target selection, and data reduction of the \texttt{LAMOST}-\textit{Kepler} project are described in \cite{2015ApJS..220...19D}. We use the effective temperature \Teff, metallicity \FeH, and surface gravity \logg of $51,385$ stars derived by \cite{2016arXiv160609149F}. The observed metallicity, \FeH, is measured as the iron abundance relative to hydrogen compared to solar in logarithmic units. After cross-matching targets observed by \textit{Kepler} (using the stellar catalog from \citealt{2014ApJS..211....2H}) and removing giants according to the prescription of \cite{2011AJ....141..108C} based on \Teff and \logg, we obtain a sample of $20,863$ main sequence stars observed by \textit{Kepler} with spectroscopic metallicities. Although \logg from \texttt{LAMOST} are less accurate than those from high-resolution spectrometry, it presents an improvement over photometric \logg used in previous occurrence rate studies. Because the source sample is predominantly magnitude-limited (\textit{Kepler} magnitude $<$ 14), the sample contains mainly G and F stars with very few cooler stars (a mean effective tempertaure $\overline{\Teff}=5990$ with a standard deviation of $590$ K). The mean metallicity of the sample is close to solar (Figure \ref{f:hist}), but the distribution is skewed with a peak around $\FeH \sim 0.25 ~\dex$ with a long tail to low metallicities ($\FeH \sim -1.5 ~\dex$) and a shorter tail to high metallicities ($\FeH \sim 0.7 ~\dex$).

\begin{figure}
	\includegraphics[width=\linewidth]{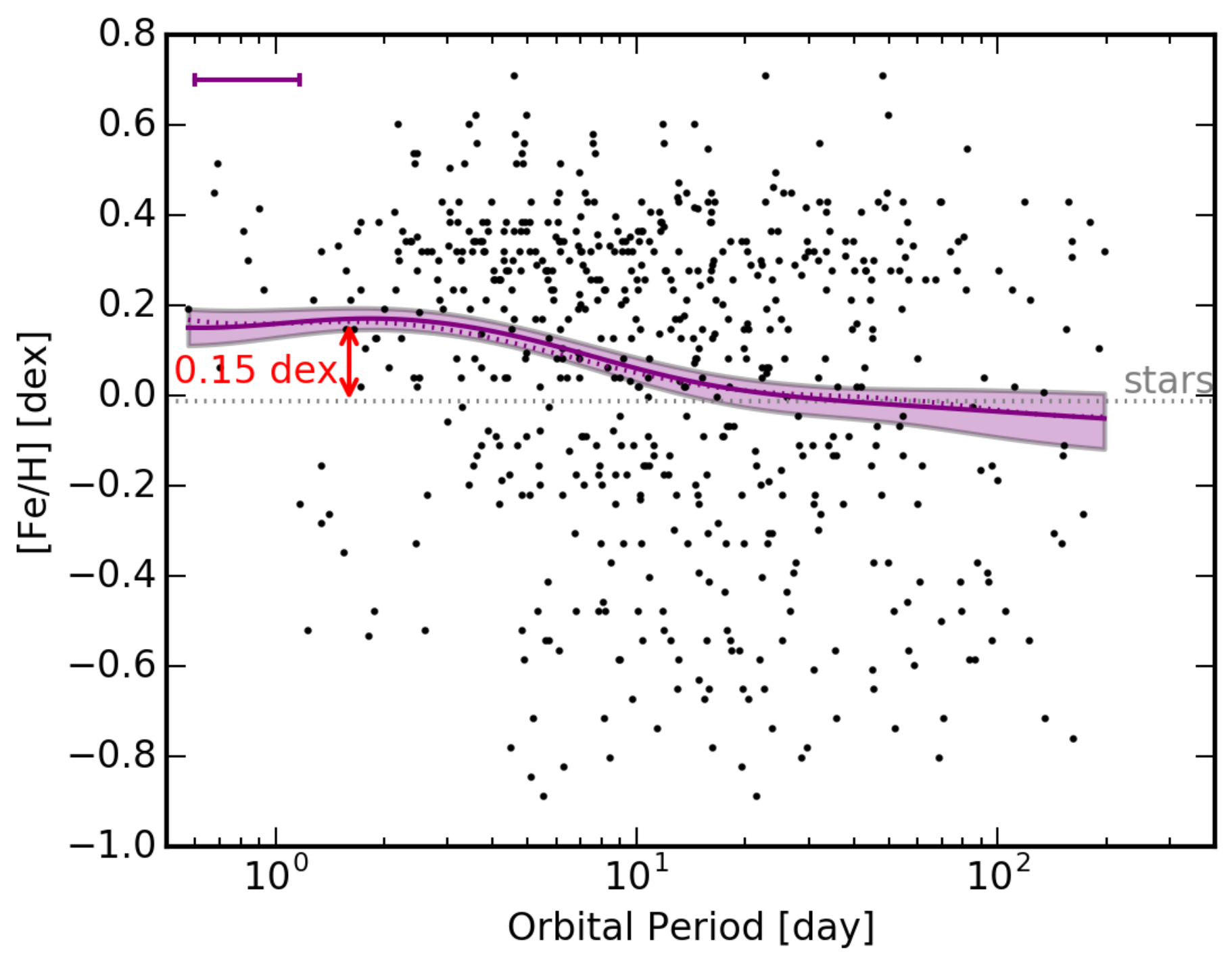}
	\caption{Host star metallicities as function of planet orbital period (black dots). The solid purple line shows  the kernel regression of the mean metallicity of the planet population (Eq. \ref{eq:mean:occ}). The shaded purple area shows the 68\% confidence interval on the mean from bootstrapping.  The kernel bandwidth of $0.29$ dex  is shown in the top left. The purple dotted line shows the mean metallicity of planet host stars (Eq. \ref{eq:mean:count}). The grey dashed line shows the mean metallicity of the stellar sample, which is consistent with solar.
	An increase with respect to longer-period planets in the host star metallicity of $\sim 0.15$ dex is evident in the planet population at orbital periods less than 10 days, indicated by the red arrow. 
	\label{f:KOIs}
	}
\end{figure}

This sample contains 665 \KOIs from the Q1-Q16 catalog \citep{2015ApJS..217...31M} after removing false positives identified by \cite{2016A&A...587A..64S}. The planet host star metallicities are displayed in Table \ref{t:occ} and shown in Figure\,\ref{f:KOIs} as a function of orbital period, together with the average stellar metallicity.

\subsection{Methods}
 We first test for a correlation between host star metallicity and the orbital periods of the planets using two nonparametric correlation tests. Spearman's rank correlation coefficient is $\rho= -0.20$ with a probability of $p=1.5e-07$ that both quantities are uncorrelated. Kendall's tau coefficient is $\tau= -0.139$ with a $p=1.2e-07$ for a lack of correlation. The host star metallicities thus show a weak but significant ($5.3\sigma$) anticorrelation with the planet orbital period. We test for the robustness of the correlation by performing a Monte Carlo simulation where we generate 10,000 sets of data where we perturb the metallicities with the typical $1-\sigma$ uncertainty of $0.2$ dex. We recover the correlation at a significance level of $4.7^{+0.5}_{-0.5}\sigma$ for both Spearman's rank and Kendall's tau.

Because the correlation is weak and the dispersion in metallicities is large, we use four different methods to investigate the metallicity-dependence of the planet population.
\begin{enumerate}
\item The average metallicity of planet host stars, \FeHcount, as in for example \cite{2012Natur.486..375B}. Because the \textit{Kepler} survey favors detection of large planets in short orbital periods around quiet stars, this metric is biased towards those planets. 
\item The average metallicity of the planet population, \FeHocc, calculated as a weighted average of the metallicity using the occurrence rate of each planet candidate, \focc, described below. Because the occurrence rates take into account transit geometry and planet detection efficiency, this metric is less biased towards the detected population and better represents the underlying exoplanet population.
\item  Non-parameteric quantile smoothing to estimate the dispersion in the host star metallicity of the planet population at different orbital periods.
\item The planet occurrence rates of a sample of stars with a super-solar metallicity (\FeH$\ge 0$ dex) and a sub-solar metallicity (\FeH$<0$ dex), which we describe in \S \ref{s:occ}.
\end{enumerate}
To calculate planet occurrence rates, we use the Q1-Q16 catalog from \cite{2015ApJS..217...31M} and associated detection completeness from \cite{2015ApJ...810...95C}. We prefer this catalog over the newer Q1-Q17 catalog from \cite{2016ApJS..224...12C} because the latter has a lower detection efficiency at longer orbital periods and is hence less complete \citep{2016arXiv160505729C}.
We calculate occurrence rates \focc for these \KOIs based on the methodology described in \cite{2015ApJ...814..130M}, using the stellar sample as defined above. The planet occurrence rates and host star metallicities are listed in Table \ref{t:occ}.

The main focus of this work is to investigate the metallicity-dependence of the exoplanet population on orbital period. Because planets form around stars with a wide rage of metallicities \cite[e.g.][]{2012Natur.486..375B}, trends in the mean metallicity are not always apparent from the raw data, e.g. Figure \ref{f:KOIs}. To visualize these trends, we use  kernel regression using the Nadaraya-Watson estimator \citep{Nada:1964,Wats:1964} to estimate how the \textit{mean} host star metallicity varies as a function of orbital period (Figure \ref{f:KOIs}, purple line). 

The  kernel regression of the mean metallicity, \FeHcount, of \KOI host stars at an orbital period, P, is given by the sum of the contributions of all $n$ \KOIs: 
\begin{equation}
	\label{eq:mean:count}
	\FeHcount(P)= \frac{\Sigma_{i=0}^{n} ~ \FeH_i ~ K(\log(P/P_i), \sigma)}{\Sigma_{i=0}^{n} ~K(\log(P/P_i), \sigma)},
\end{equation}
where $\FeH_i$ and $P_i$ are the observed metallicity and orbital period for each planet candidate as reported in Table \ref{t:occ}. We use a log-normal kernel 
\begin{equation}
K(\log P, \sigma)= \frac{1}{\sqrt{2\pi}\sigma} e^{-0.5 (\log P/\sigma)^2}
\end{equation}
with a  constant bandwidth $\sigma$.  We estimate a bandwidth of $\sigma=0.29$ using maximum likelihood cross-validation.

The  kernel regression of the mean metallicity of the \textit{exoplanet population} is calculated by weighing the contribution of every \KOI with its occurrence rate, \focc:
\begin{equation}
	\label{eq:mean:occ}
	\FeHocc(P)= \frac{\Sigma_{i=0}^{n} ~\FeH_i ~f_{\rm occ, i} ~K(\log(P/P_i), \sigma)}{\Sigma_{i=0}^{n} ~f_{\rm occ, i} ~K(\log(P/P_i), \sigma)}.
\end{equation}

Confidence intervals ($1-\sigma$) of the \textit{mean} metallicity are calculated  using a bootstrapping method, which goes as follows. First, we generate 10,000 bootstrapped samples from the original sample. Each bootstrap is a random draw with replacement from the original sample with a draw size equal to the original sample size. Second, we calculate the kernel regression of each bootstrapped sample using equation \ref{eq:mean:occ}. Third, we calculate the 68th percentiles at each period from the kernel regression of all bootstrapped samples. 

\begin{figure}
	\includegraphics[width=\linewidth]{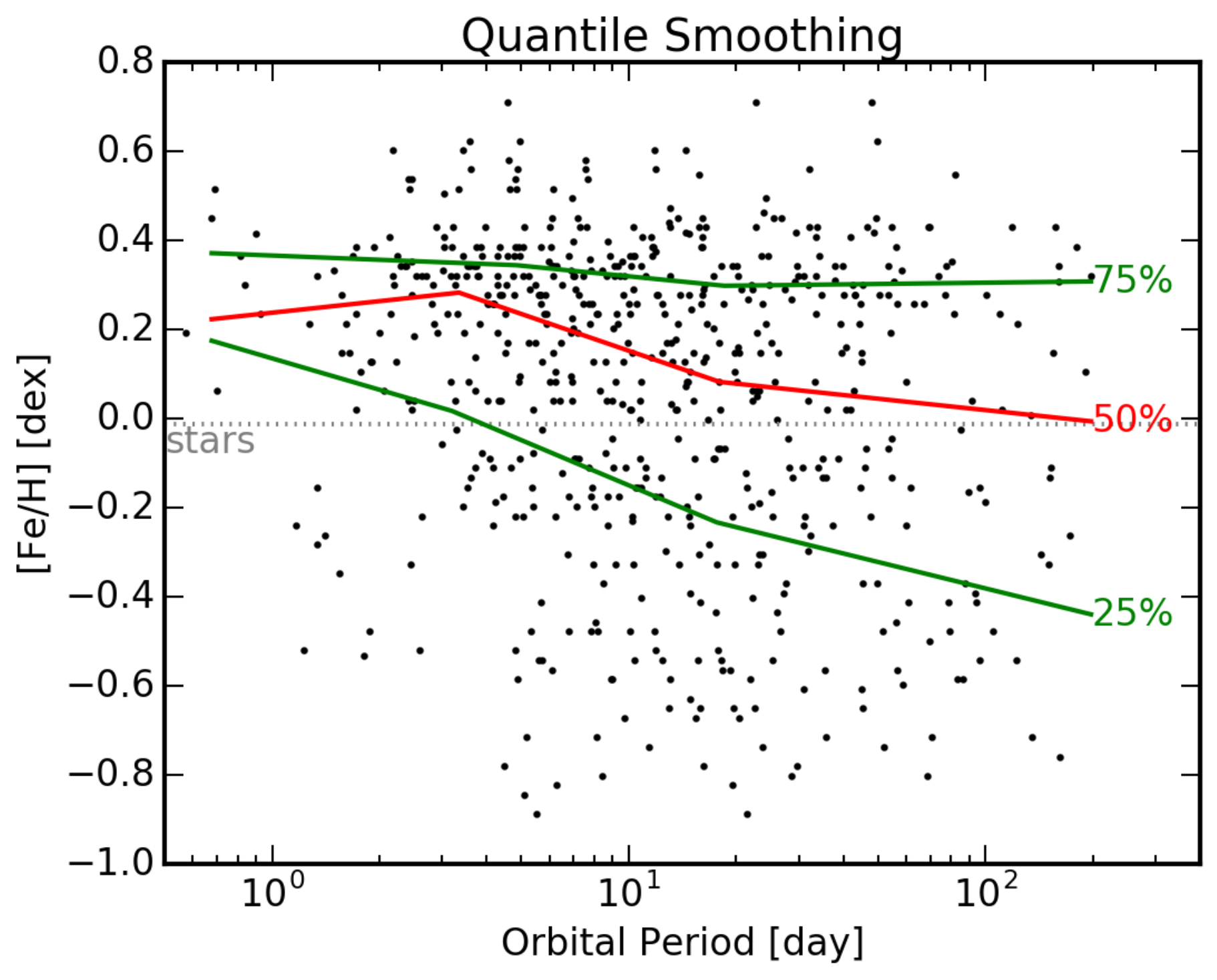}
	\caption{Non-parametric quantile smoothing of the 25\%, 50\%, and 75\% quantiles (solid lines) of the host star metallicities of the planet population as a function of orbital period. Individual datapoint are shown with black dots. The grey dashed line shows the mean metallicity of the stellar sample, which is consistent with solar.
	The increase in metallicity at short orbital periods becomes more prominent towards the lower quartiles. The dispersion in the metallicities increases towards larger orbital period.
	\label{f:quant}
	}
\end{figure}

The $1-\sigma$ confidence intervals reflect the uncertainty in the mean metallicity, not the intrinsic scatter around the mean.  An estimate of the dispersion in the data is given by non-parameteric quantile smoothing \citep{KNP.94, RSSB:RSSB437}. We calculate the 25\%, 50\% and 75\% quantiles of the metallicity of the planet population which are shown in Figure \ref{f:quant}.

\begin{figure}
	\includegraphics[width=\linewidth]{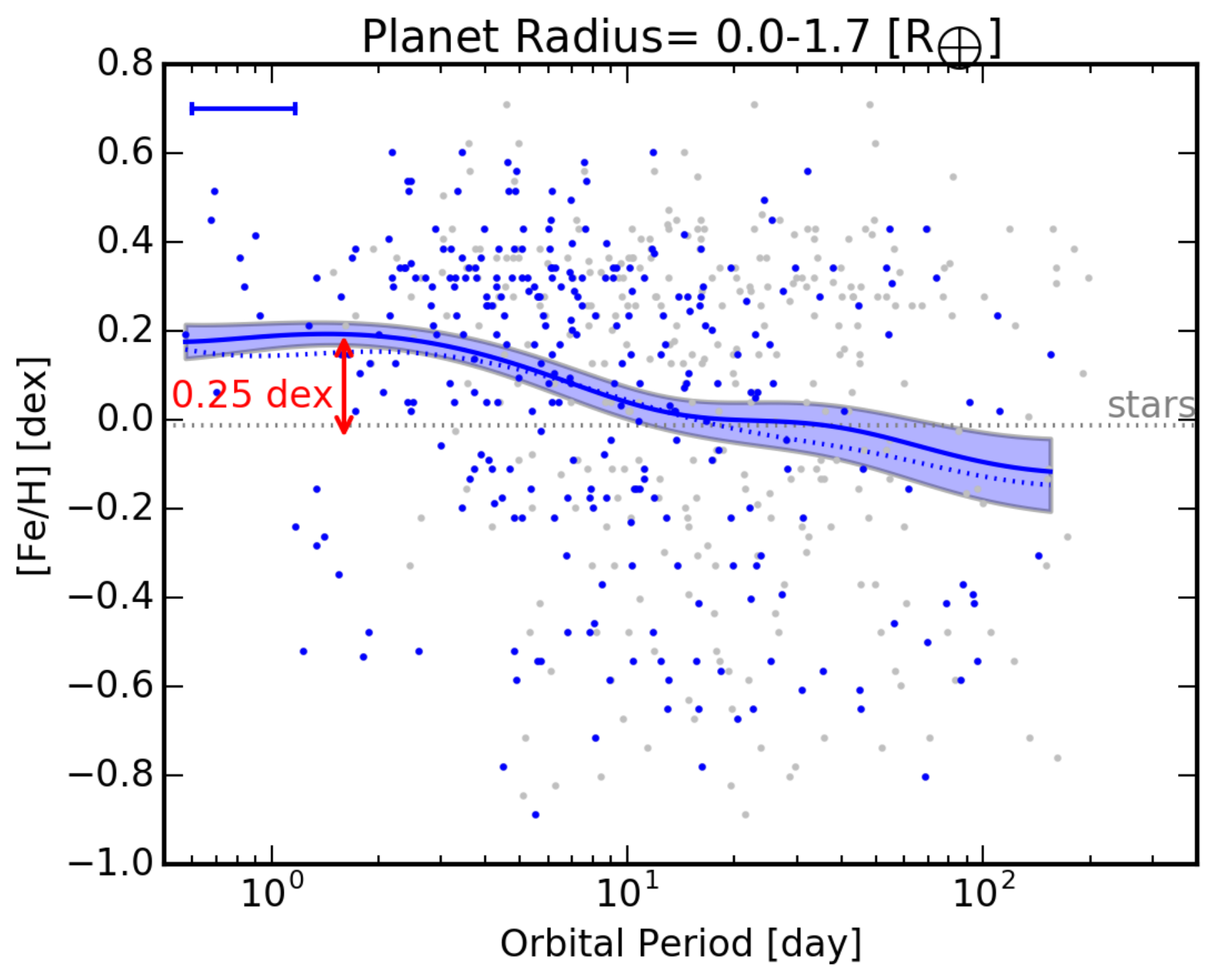}
	\includegraphics[width=\linewidth]{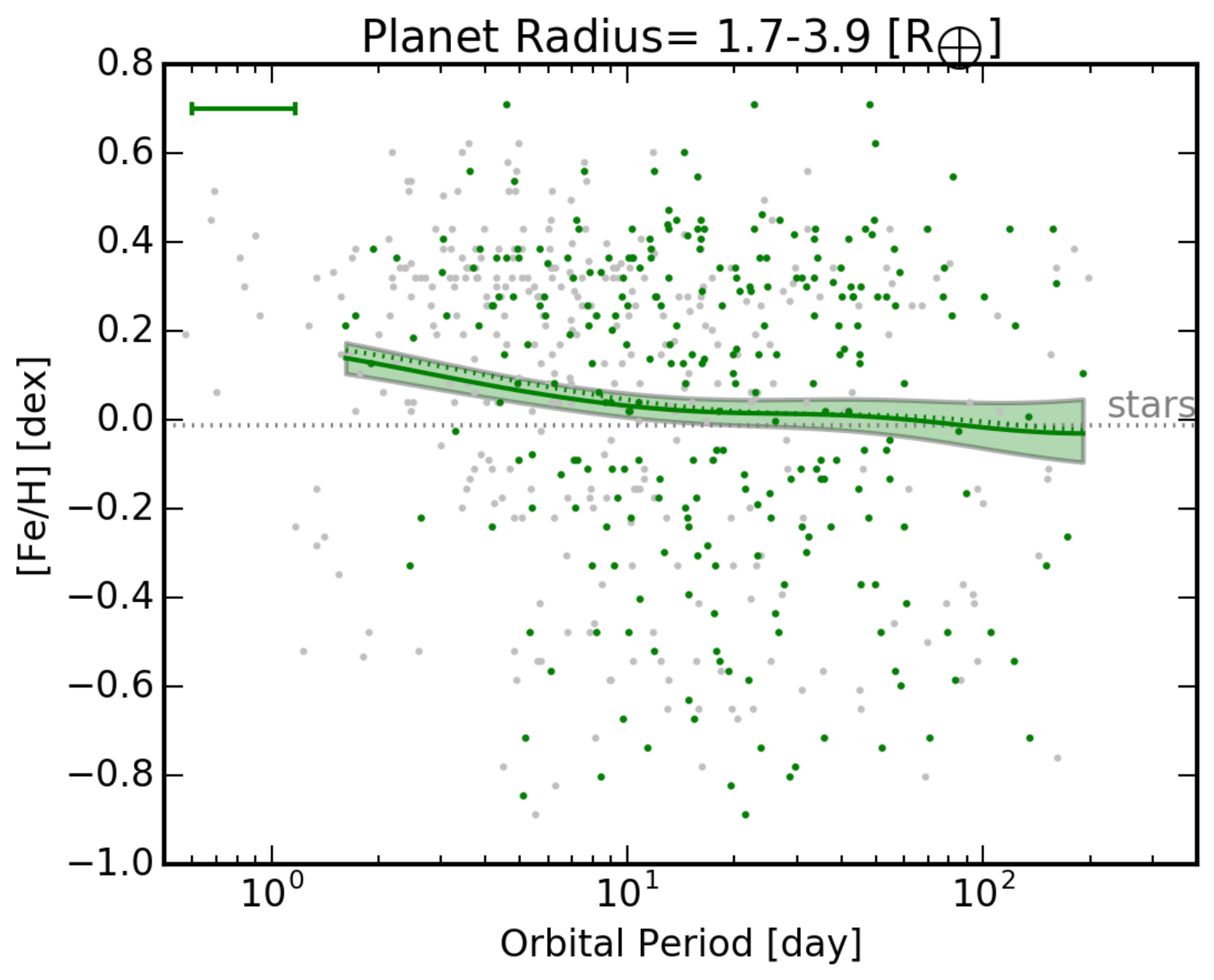}
	\includegraphics[width=\linewidth]{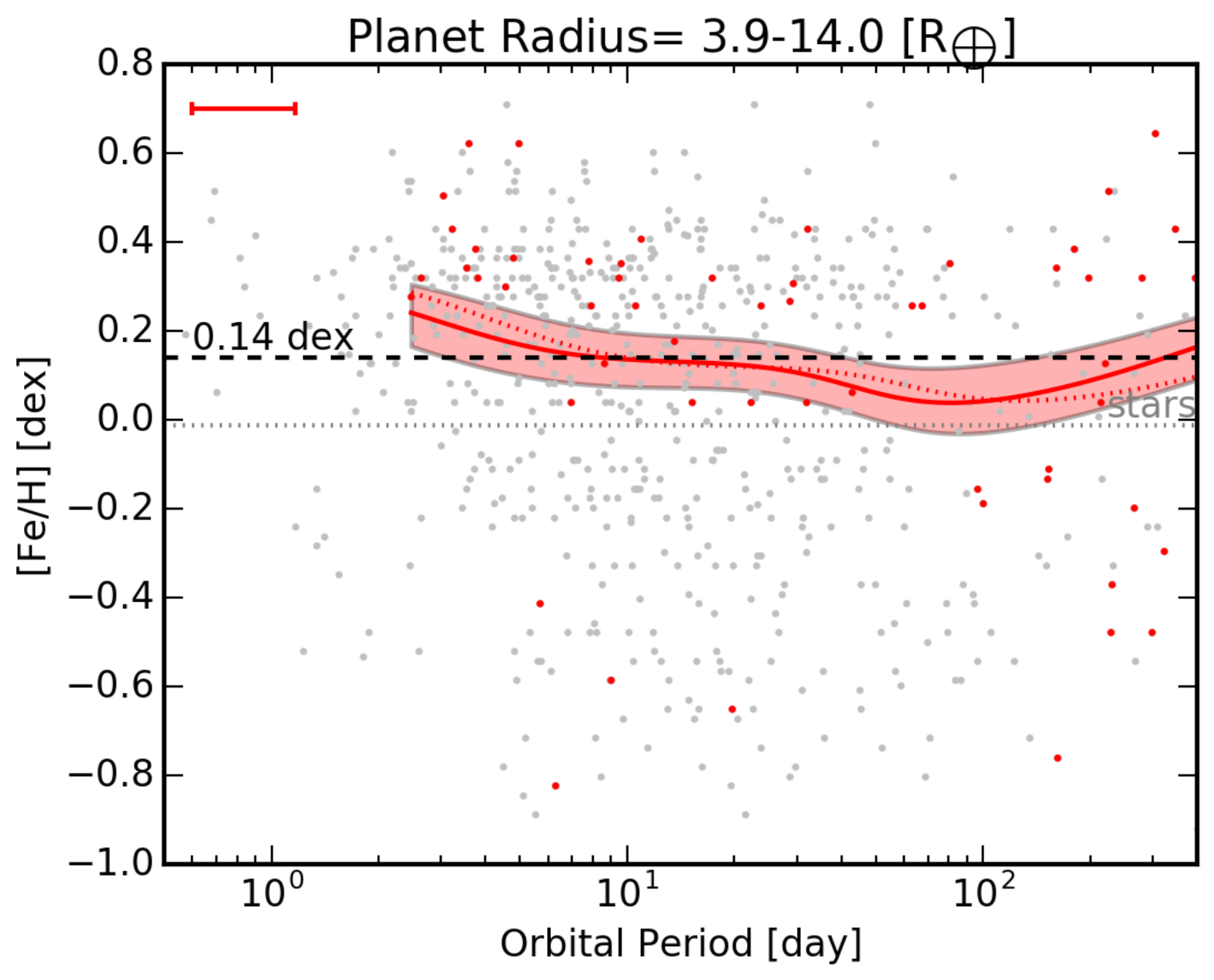}
	\caption{Same as Figure \ref{f:KOIs} for, from top to bottom: small planets ($R_p=[0,1.7] ~R_\oplus$), intermediate-sized planets ($R_p=[1.7, 3.9] ~R_\oplus$), and large planets ($R_p>[3.9] ~R_\oplus$). The difference in host star metallicity is most pronounced in the smallest planets, highlighted by the red arrow. The metallicity of giant planets is consistent with a period-independent mean metallicity of $0.14$ dex, indicated by the black dashed line. 
	\label{f:sizes}
	}
\end{figure}

\section{Results}\label{s:results}
\subsection{Metallicity vs. Orbital Period}
The average metallicity of the planet population increases interior to an orbital period of $\sim{} 10$ days by about $0.15$ dex (Fig. \ref{f:KOIs}). We note that this metallicity is similar to that of giant planet hosts in the sample: The host star metallicity of the 44 planets larger than $4~R_\oplus$ is $\FeH=0.14 \pm 0.04 ~\dex$, consistent with previous estimates for \textit{Kepler} giant planet hosts of $\FeH=0.15 \pm 0.03 ~\dex$ \citep{2012Natur.486..375B} and $\FeH=0.18 \pm 0.02 ~\dex$ \citep{2014Natur.509..593B}. Because the scatter around the mean metallicity is non-Gaussian, we perform two statistical tests to asses the significance of the difference in host star metallicity inside and outside of this orbital period. First, the Mann-Whitney test computes the probability, $p_{\rm MW}$, that the two distributions have the same mean. Second, the Kolmogorov-Smirnoff test computes the probability, $p_{\rm KS}$, that the two distributions are drawn from the same parent distribution. We find probabilities of $p_{\rm MW}=4.5 \cdot 10^{-6}$ [ 4.6 $\sigma$] and $p_{\rm KS}=2.1 \cdot 10^{-5}$ [4.3 $\sigma$] that the metallicity interior and exterior to a 10-day period have the same mean, or are drawn from the same distribution, respectively.  The dispersion in the metallicity of the planet population increases with orbital period (Fig. \ref{f:quant}). The 50\% quantile of the metallicity shows an increase at short orbital periods which is similar in magnitude to that of the average metallicity. The increase is smaller for the 75\% quantile and much larger for the 25\% quantile.

The  increase in metallicity at short orbital periods is mainly driven by the smallest planets in the sample (Fig. \ref{f:sizes}). Throughout this work, we adopt the same planet size ranges as \cite{2014Natur.509..593B} for rocky planets ($R_P < 1.7 ~R_\oplus$), mini-Neptunes ($R_P=1.7-3.9 ~R_\oplus$), and giant planets ($R_P>3.9 ~R_\oplus$). Although the choice of planet radii are somewhat arbitrary, they do reflect the boundary between rocky planets and planets with a gaseous envelope at $~1.6 R_\oplus$ \citep{2015ApJ...801...41R}, and between planets whose mass is dominated by their rocky cores versus by a gaseous envelope \citep{2014ApJ...792....1L}. We have verified that our results are not sensitive to the exact choice of the planet radius boundaries. The increased metallicity at short orbital periods is most significant for rocky planets ($p_{\rm MW}=3.8 \cdot 10^{-7}$ [5.1 $\sigma$], $p_{\rm KS}=5.0 \cdot 10^{-5}$ [4.1 $\sigma$]). The Mann-Whitney and Kolmogorov-Smirnoff tests are inconclusive for mini-Neptunes ($p_{\rm MW}=0.12$, $p_{\rm KS}=0.2$) and for gas giants ($p_{\rm MW}=0.04$, $p_{\rm KS}=0.1$).  We note that the significance of the correlation is higher in the rocky planet sample than in the entire sample. This indicates that the lower significance of the correlation in the mini-Neptune and gas giant samples is due to a correlation that is intrinsically weaker and not due to low-number statistics.

\begin{figure}
	\includegraphics[width=\linewidth]{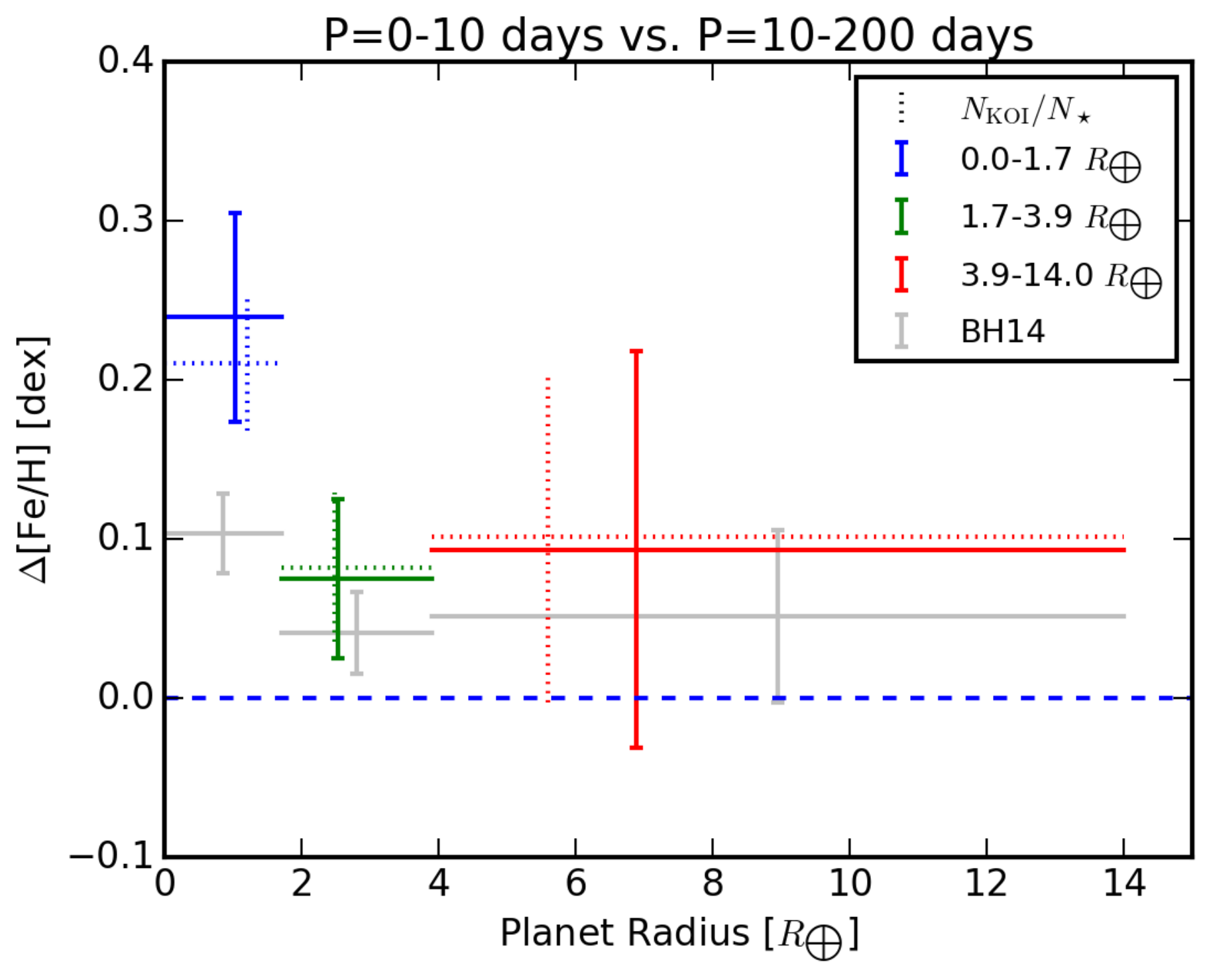}
	\caption{Difference in metallicity between planet at short ($P<10 ~\rm{day}$) and long ($P>10 ~\rm{day}$) orbital periods. Smaller planets show a stronger metallicity-dependence. Solid colors show the occurrence-weighted metallicities (Eq. \ref{eq:bin:occ}), dotted colors show the average metallicities without taking into account planet occurrence rates. The gray colors show the trend of higher host star metallicity at shorter orbital periods is also present in the dataset from \cite{2014Natur.509..593B}.
	\label{f:diff}
	}
\end{figure}

Because the Mann-Whitney test and Kolmogorov-Smirnoff test do not take into account planet occurrence rates, we also calculate a different statistic, $\Delta\FeH$, to compare the mean metallicity of the planet population interior and exterior to 10 days. First, we calculate the mean host star metallicity of the planet population in a given period and radius range bin $\{R_P,P\}$ as:
\begin{equation}\label{eq:bin:occ}
\FeHocc\{R_P,P\}= \frac{\Sigma_i^{\{R_P,P\}} ~f_{{\rm occ},i} ~ \FeH_i}{\Sigma_i^{\{R_P,P\}} ~f_{{\rm occ}, i}}.
\end{equation}
Confidence intervals at $1-\sigma$ are once again calculated by bootstrapping the sample 10,000 times, re-calculating the occurrence-weighted metallicity for each draw, and taking the 68th percentile from the distribution of all draws.
The difference between the occurrence-weighted metallicity interior and exterior to a 10 day orbital period, \DeltaFeH, is defined as:
\begin{equation}\label{eq:delta}
\begin{split}
\DeltaFeH= &\FeHocc\{R_P,P<10~d\} \\
&- \FeHocc\{R_P,P\ge 10~d\}
\end{split}
\end{equation}
and shown in Figure \ref{f:diff} for the same planet size ranges as before. Confidence intervals at $1-\sigma$ are estimated by quadratically adding the confidence intervals on $\FeHocc\{R_P,P\}$.
 
The difference in host star metallicity of the planet population is largest ($\DeltaFeH=0.25\pm0.07$ dex) for rocky planets, smaller for mini-Neptunes ($\DeltaFeH=0.08\pm0.05$ dex), and not significant for giant planets ($\DeltaFeH=0.10\pm0.12$ dex). We note that the same trend is present in the data-set from \cite{2014Natur.509..593B}, albeit at lower amplitude. The difference may be due to a different sample selection, or to the calibration of the \texttt{LAMOST} metallicities, which we will discuss in Appendix \ref{s:cal}.

\begin{figure}
	\includegraphics[width=\linewidth]{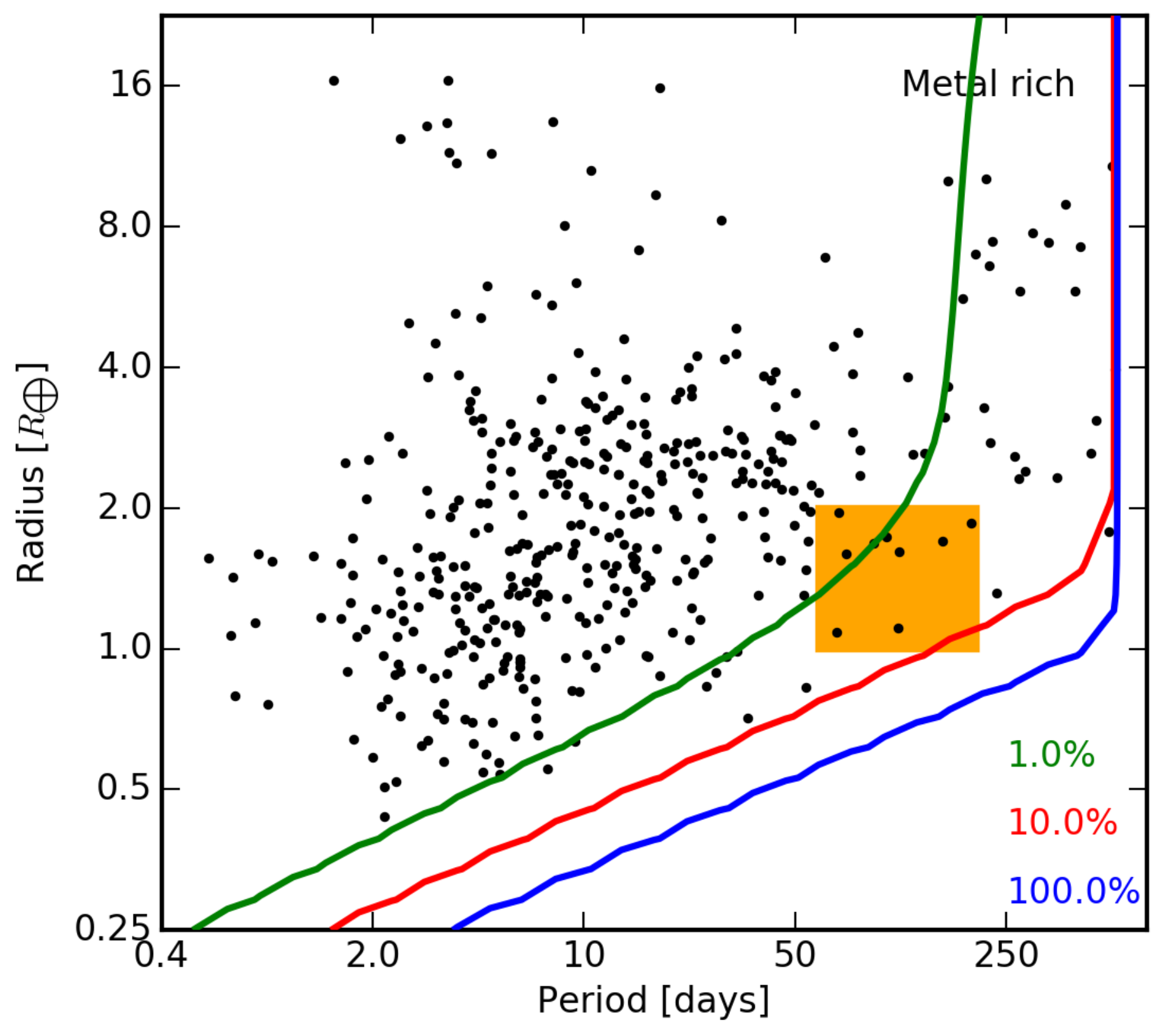}
	\includegraphics[width=\linewidth]{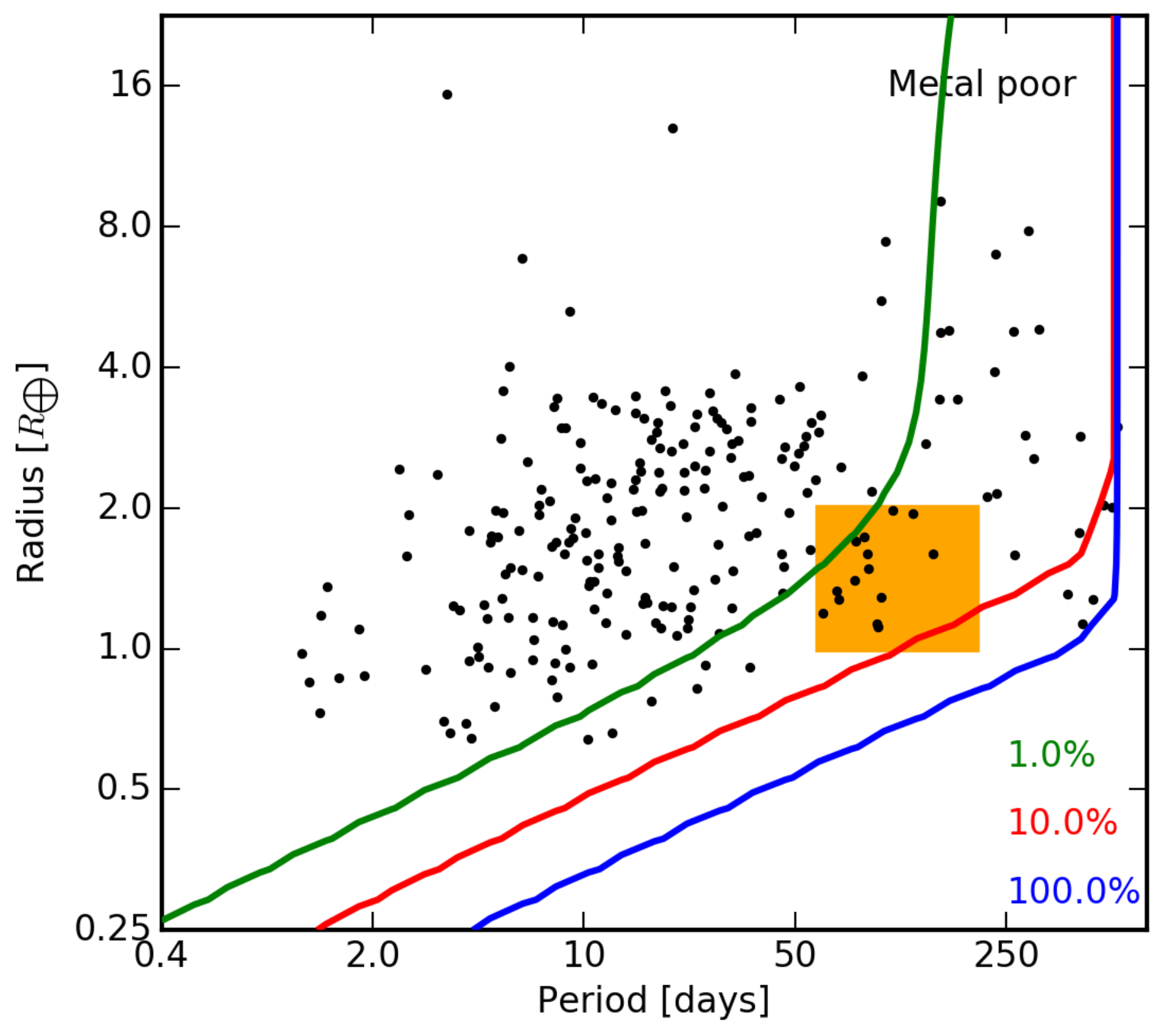}
	\caption{Completeness contours and \KOIs for the super-solar and sub-solar metallicity sample. The orange box indicates the region identified by \cite{2016OLEB..tmp...30A} as containing no planets in the sub-solar sample.
	\label{f:completeness}
	}
\end{figure}

\begin{figure}
	\includegraphics[width=\linewidth]{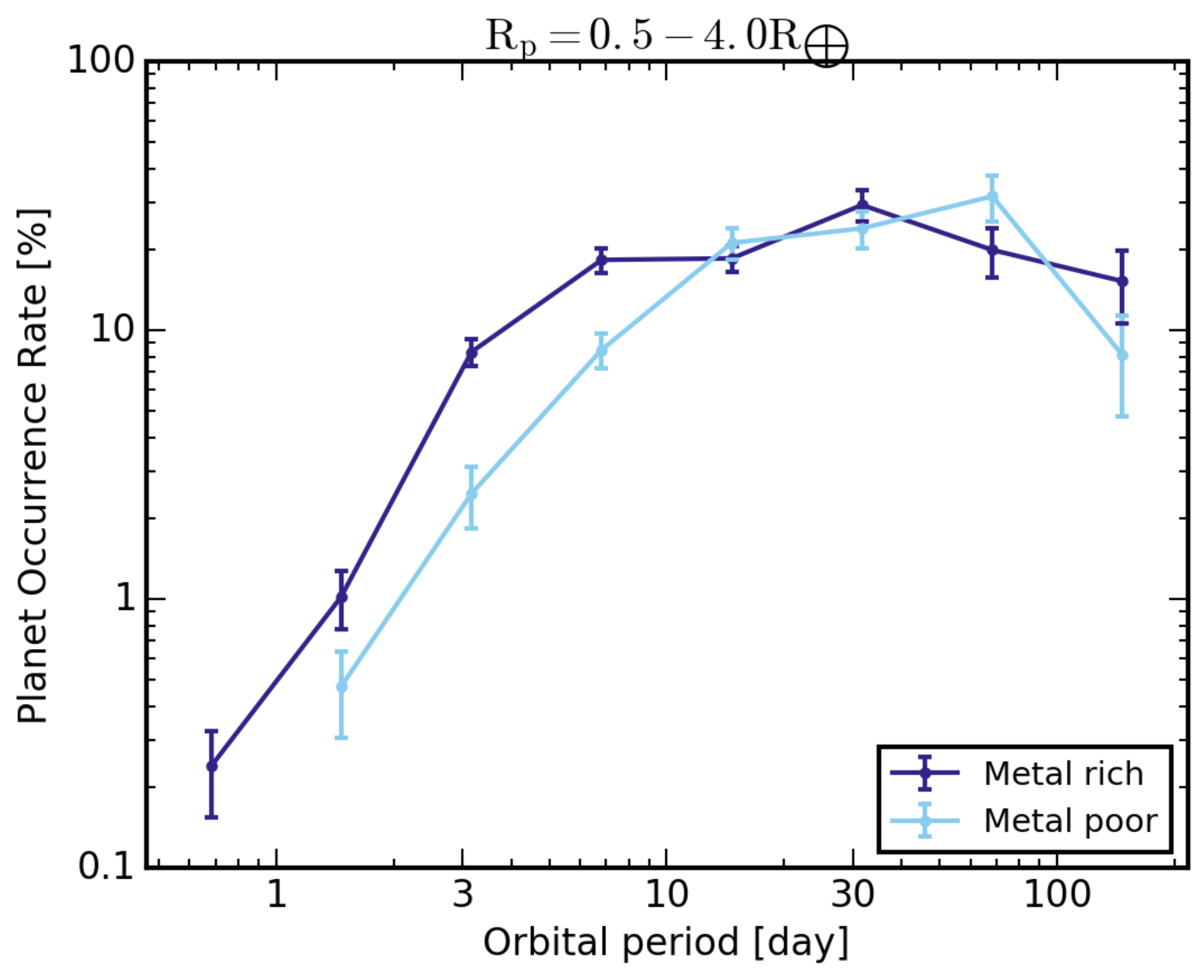}
	\caption{Planet occurrence as a function of orbital period for super-solar and sub-solar metallicity stars. The planet occurrence rate interior to the cutoff at $P\sim10 ~\rm{day}$ is three times higher for stars with super-solar metallicity. At larger orbital periods, the occurrence rates are similar within errors.
	\label{f:radial}
	}
\end{figure}

\subsection{Planet Occurrence}\label{s:occ}
Although we find a large difference in host star metallicity for planets at short orbital periods, this is also the region where planet occurrence rates are low ($\sim{} 20\%$) compared to longer orbital periods ($\gtrsim{} 90\%$). To asses how the metallicity dependence of the short period planets affects the overall planet population, we calculate planet occurrence rate for a subsample of sub-solar ($\FeH<0$) and super-solar ($\FeH \ge 0$) metallicities. The stars in both samples have similar levels of photometric noise (CDPP, \citealt{2012PASP..124.1279C}), and the detection efficiencies as a function of planet radius and orbital period are very similar (see Fig. \ref{f:completeness}). Figure \ref{f:radial} shows the occurrence rates of both samples as function of orbital period for planets smaller than 4 $R_\oplus$. The inclusion of giant planets or the exclusion of mini-Neptunes does not significantly change the trend with metallicity. The super-solar sample has an almost three times higher occurrence rate within a 10 day orbital period of $29.6 \pm 2.0 \%$ compared to the sub-solar sample with a rate of $11.9 \pm 1.4 \%$. This trend is consistent with the higher mean metallicity for planet hosts identified in the previous section. At longer orbital periods, where the bulk of the planet population resides ($88.5 \pm 4.9 \%$ between 10 and 200 days), there is no statistically significant difference between the planet occurrence rates for super-solar ($89.0 \pm 6.5 \%$) and sub-solar ($89.9 \pm 7.6 \%$) metallicity stars.

\subsection{Trends toward the Habitable Zone}\label{s:HZ}
\cite{2016OLEB..tmp...30A} identify a lack of planets smaller than $2 ~R_\oplus$ towards the habitable zone around stars with a super-solar metallicity by looking at confirmed planets in the \cite{2014Natur.509..593B} sample. This region, corresponding to an orbital period range of 60-200 days, is highlighted with the orange box in Figure \ref{f:completeness}. Our sample contains \KOIs in this box for both samples. The integrated planet occurrence rates are twice as high for stars of sub-solar metallicity ($26.7 \pm 7.1 \%$) compared to stars of super-solar metallicity ($13.8 \pm 4.6 \%$).  This trend is, however, not statistically significancant at $1.5 ~\sigma$ and a larger sample of stars with spectroscopic metallicities is required to  refute or confirm its existence. 

\section{Discussion}\label{s:D}
We will now discuss two different scenarios that may explain the observed trend. The super-solar metallicity of hot rocky planet hosts is similar to that of giant planets, while their location ($P<10$ days) coincides with that of hot Jupiters. This may indicate that hot Jupiters and hot rocky planets have a common formation mechanism that sets them aside from colder ($P>10$ days) rocky planets and mini-Neptunes. Hot rocky planets may represent a population of planets that formed like hot Jupiters but did not enter runaway gas accretion. They may have formed with a gaseous envelope, but were not massive enough to prevent their gaseous envelopes from escaping, reducing their sizes to those of rocky planets. This picture is consistent with the observed lack of hot Neptunes at very short orbital periods ($P \lesssim 2.5$ days) that can be explained by stripping of their gaseous envelopes \citep{2016NatCo...711201L,2016A&A...589A..75M}. However, while hot-Jupiters are typically not found in multi-planet systems \citep{2012PNAS..109.7982S}, hot rocky planets are. The increased host star metallicity for hot rocky planets does not disappear if we consider only observed multi-planet systems ($p_{\rm MW}=8.7 \cdot 10^{-5}$ [3.9 $\sigma$], $p_{\rm KS}=5.3 \cdot 10^{-3}$ [2.8 $\sigma$]).

An alternative explanation is that the regions where planets form or halt their migration extend closer-in around metal-rich stars. If we interpret the drop in planet occurrence rates interior to a ten-day orbital period as a signature of the protoplanetary disk inner edge \cite[e.g.][]{2015ApJ...798..112M}, this edge must be closer in around metal-rich stars. Figure \ref{f:radial} shows that the orbital period of the inner disk edge around metal-rich stars must be half that around metal-poor stars. This corresponds to a difference in semi-major axis of a factor 1.6. We note that the expected metallicity difference between M stars and sun-like stars in the \textit{Kepler} sample is less than $0.1$ dex \citep[e.g.][]{2012ApJS..201...15H}, and is too small to influence the stellar-mass dependent trends identified in \cite{2015ApJ...798..112M}.

The inner edge of the dust disk may serve as a planet trap \citep[e.g.][]{2016A&A...590A..60B} or a preferred site of planet formation \citep[e.g.][]{2014ApJ...792L..27B}. The exact scaling-law between the location of the dust disk inner edge and disk metallicity is not clear. A high disk metallicity leads to a larger dust-to-gas ratio, increasing the disk opacity, and moving the dust sublimation front inward \citep[e.g.][eq. 3]{2009A&A...506.1199K}. However, the complex interplay between dust opacity and disk structure at the sublimation front makes it hard to estimate a scaling law for the inner disk edge with disk metallicity. Self-consistent radiation hydro-dynamical modeling of the sublimation front, such as that in \citep{2016ApJ...827..144F}, around pre-main sequence sun-like stars will be necessary to pin down the scaling law between the dust disk inner edge and the stellar metallicity. 

The inner edge of the gas disk is a strong trap for migrating rocky planets \citep[e.g.][]{2007ApJ...654.1110T}. How the inner edge of the gas disk depends on metallicity is less clear. The location of the inner edge can be calculated from the balance between the magnetic pressure from the stellar magnetic field and the ram pressure from the ionized gas. A higher metallicity increases the mean molecular weight by only a small factor, and the corresponding increase in ram pressure is not sufficient to move the inner disk edge inward. How the stellar magnetic field of pre-main-sequence stars depends on their metallicity is not clear. A weaker magnetic field around metal-rich stars is required to move the inner edge inward. Alternatively, if the inner disk edge is determined from the gas co-rotation radius as in \cite{2015ApJ...798..112M}, high-metallicity pre-main-sequence stars must be faster rotators to explain the observed trend. There does not seem to be any correlation between rotation and metallicity, at least for M dwarfs \citep{2016ApJ...821...93N}.

\section{Conclusion}\label{s:C}
We have characterized the orbital-period dependence of the \textit{Kepler} exoplanet population using $>20,000$ medium-resolution spectroscopic metallicities from the \texttt{LAMOST} survey for main-sequence G and F stars. For the first time we are able to calculate planet occurrence rates for the \textit{Kepler} sample based on spectroscopic metallicities. We find that:
\begin{itemize}
\item The metallicities of the stellar sample is consistent with solar, while giant planets have an increased host star metallicity of $0.14\pm0.04$ dex.
\item The exoplanet population, which is dominated by planets smaller than $\sim 4$ Earth radii, shows an increased host star metallicity of [Fe/H]$\simeq{} 0.15\pm0.05$ dex) interior to a 10-day orbital period. At longer orbital periods metallicities are consistent with solar.
\item The super-solar metallicity at short orbital periods is most significant for rocky planets ($R_P<1.7 ~R_\oplus$), where metallicity differs by [Fe/H]$\simeq 0.25\pm0.07$ dex. The difference in metallicity for hot mini-Neptunes ($R_P=1.7-3.9 ~R_\oplus$) is smaller and less significant at [Fe/H]$\simeq 0.08\pm0.05$ dex. Hot gas-giants ($R_P>3.9 ~R_\oplus$) do not show a significant metallicity variation with orbital period ([Fe/H]$\simeq 0.1\pm0.12$ dex).
\item The occurrence rate of planets interior to a 10-day orbital period is almost three times higher for super-solar metallicity stars ([Fe/H]$\ge 0$, $29.6 \pm 2.0 \%$) than for stars with a sub-solar metallicity ([Fe/H]$<0$ , $11.9 \pm 1.4 \%$). Exterior to 10-day orbital periods, there is no significant difference between planet occurrence rates around stars of super-solar metallicity ($89.0 \pm 6.5 \%$) and sub-solar metallicity ($89.9 \pm 7.6 \%$).
\item The increased host star metallicity of hot rocky planets suggests they may share a formation mechanism with hot-Jupiters that is distinct from the population of rocky planets and mini-Neptunes at orbital periods that shows no metallicity dependence. Alternatively, planet formation regions may extend closer-in around stars with higher metallicity, which is supported by hot rocky planets also appearing in multi-planets systems, in contrast to hot Jupiters that are typically single.
\item We  do not find statistically significant evidence for a trend previously identified by \cite{2016OLEB..tmp...30A} that small planets toward the habitable zone are preferentially found around low-metallicity stars.  Although the occurrence rates of planets smaller than 2 Earth radii with orbital periods between 70 and 200 days are higher for stars with a sub-solar metallicity ($26.7 \pm 7.1 \%$) compared to stars of super-solar metallicity ($13.8 \pm 4.6 \%$)  , this difference is only $1.5\sigma$.
\end{itemize}

Although the \textit{Kepler} spacecraft has finished its main mission, ongoing characterization of the stellar content of the \textit{Kepler} field with surveys like \texttt{LAMOST} and in the future \texttt{GAIA} can shed new light on the planet formation process. In particular trends for exoplanets in the habitable zone, where the detection efficiency is low,  can be refuted or placed on a strong statistical footing with a larger dataset of spectroscopic metallicities.

\vspace{1em}{\it Acknowledgments:}\\

This paper includes data collected by the \textit{Kepler} mission. Funding for the \textit{Kepler} mission is provided by the NASA Science Mission directorate.
We thank the referee and the statistical editor for a constructive review that has improved the quality of the paper.
The authors thank Mario Flock and Aline Vidotto for helpful discussions on the disk inner edge. We would also like to thank Dean Billheimer and his students for statistical advice through the Statistical Consulting course.
JM-\.Z acknowledges the grant number NCN 2014/13/B/ST9/00902.
This material is based upon work supported by the National Aeronautics and Space Administration under Agreement No. NNX15AD94G for the program “Earths in Other Solar Systems”. 
The results reported herein benefited from collaborations and/or information exchange within NASA’s Nexus for Exoplanet System Science (NExSS) research coordination network sponsored by NASA’s Science Mission Directorate.

\ifastroph
	\bibliography{paper.bbl}
\else	
	\bibliography{/Users/mulders/Dropbox/papers/papers3,/Users/mulders/Dropbox/papers/books}
\fi

\appendix

\begin{figure}
	\includegraphics[width=\linewidth]{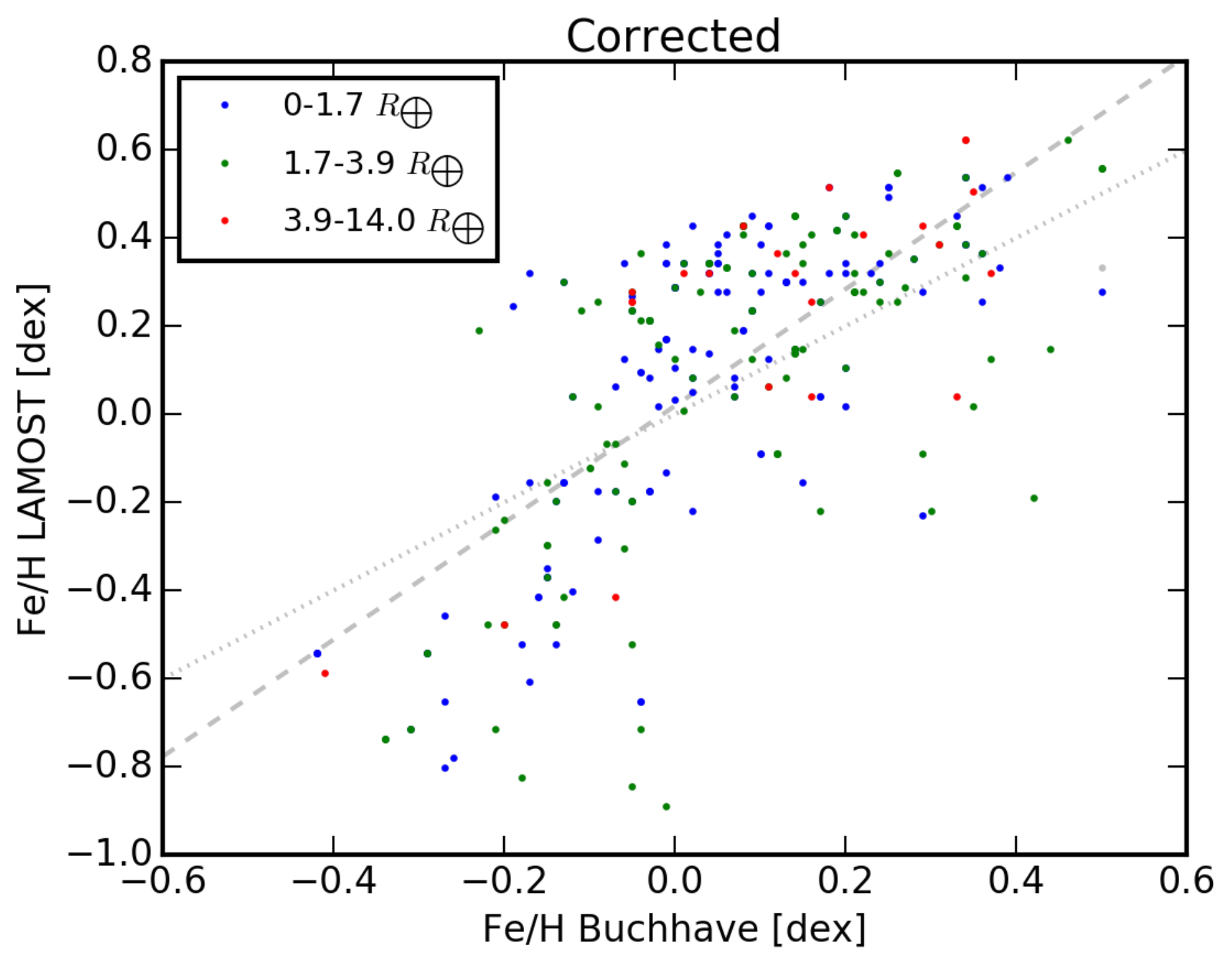}
	\caption{Comparison of host star metallicities for planet candidates present in both the \cite{2016arXiv160609149F} and \cite{2014Natur.509..593B} sample. The dotted line shows a 1:1 correlation. The \texttt{LAMOST} metallicities span a range that is 30\% larger (dashed line) than those of  \cite{2014Natur.509..593B}.
	\label{f:compare}
	}
\end{figure}

\section{Metallicity scaling}\label{s:cal}
The stellar metallicities from \cite{2016arXiv160609149F} used in this paper span a range \FeH$\simeq$\,[$-0.8$,$0.6$] that is wider than the range of \FeH$\simeq$\,[$-0.5$,$0.5$] for \textit{Kepler} planet hosts in \cite{2014Natur.509..593B}. \cite{2016arXiv160609149F} note that there is a systematic trend in the metallicities derived from \texttt{LAMOST} data compared to literature values, which is ``the result of both the low resolution of the \texttt{LAMOST} spectra and the non-uniform distribution of templates in the parameters space.'' The authors propose a correction to the measured metallicities from their pipeline, based on a comparison with the high-resolution spectroscopic metallicities from the \texttt{Apokasc} catalog of red giant stars. Throughout this paper, we have used these corrected metallicities.

However, the correction for main-sequence stars may not necessarily be the same as that for red giant stars, and may explain the wider range in metallicities in this work compared to \cite{2014Natur.509..593B}. Figure \ref{f:compare} shows the metallicities used in this paper compared to those in \cite{2014Natur.509..593B}, for the 299 planets that are present in both samples. A linear fit to the data (dashed line) shows that the corrected metallicities are on average 30\% larger than the metallicities from \cite{2014Natur.509..593B}. There is no trend between metallicity and planet size that could influence the results of this paper, though a scatter around the mean is present  with a median deviation of $0.20$ dex. However, \cite{2016arXiv160609149F} do not find a significant offset between corrected metallicities and literature objects, which are mostly main-sequence stars.

To asses the impact of a different metallicity correction on our results, we repeated the analysis of Section \ref{s:results} with a metallicity correction that reproduces the mean metallicity for the main-sequence stars as described above. Using this new correction, the metallicity increase at short orbital periods is $\FeH=0.18\pm0.05$, which is $\sim{}30\%$ smaller. This indicates that the systematic uncertainty in the derived trend is of the same order as the bootstrapping uncertainty.  
Note that the significance of this result does not change, as the $1-\sigma$ confidence intervals scale accordingly. The Mann-Whitney and Kolmogorov-Smirnoff tests produce identical results. Similarly, the planet occurrence rates derived in Section \ref{s:occ} do not change as a different metallicity correction does not influence which stars are in the metal-rich and metal-poor sample.

\end{document}